\documentclass[preprint,12pt]{elsarticle}

\usepackage{graphicx}

\usepackage{algorithm}
\usepackage[noend]{algpseudocode}

\usepackage{amsmath}
\usepackage{amssymb}

\usepackage{lineno}

\journal{Physica A}

\begin{document}

\begin{frontmatter}


\title{Spectral density of equitable core-periphery graphs}



\author{Paolo Barucca}

\address{University College London}

\begin{abstract}
Core-periphery structure is an emerging property of a wide range of complex systems and indicate the presence of group of actors in the system with an higher number of connections among them and a lower number of connections with a sparsely connected periphery. 
The dynamics of a complex system which is interacting on a given graph structure is strictly connected with the spectral properties of the graph itself, nevertheless it is generally extremely hard to obtain analytic results which will hold for arbitrary large systems.
Recently a statistical ensemble of random graphs with a regular block structure, i.e. the ensemble of equitable graphs, has been introduced and analytic results have been derived in the computationally-hard context of graph partitioning and community detection. \\
In this paper, we present a general analytic result for a ensemble of equitable core-periphery graphs, yielding a new explicit formula for the spectral density of networks with core-periphery structure. 

\end{abstract}


\end{frontmatter}


\section*{Introduction}
From stability analysis to noise reduction, spectral theory of random matrices and random graphs is crucial to predict the stability and the dynamics of systems with large number of interacting components.\\
Core-periphery structure, in which an high-density core is less densely connected to a low-density periphery, has been documented in a large variety of complex systems, such in social networks \cite{everett1999centrality,boyd2006computing,borgatti2000models}, world trade networks \cite{smith1992structure}, and financial networks\cite{fricke2015core,barucca2016disentangling}. \\
It has been investigated with various metrics, as for instance MINRES \cite{boyd2006computing}, non-backtracking centrality \cite{martin2014localization}, probability marginals \cite{zhang2015identification} and simply, but efficiently, measuring degree centrality \cite{rombach2014core,barucca2016centrality}; recently PageRank has been shown to be the most robust measure of coreness in heterogenous degree-corrected SBM graphs \cite{barucca2016centrality}. \\
In this work, we study core-periphery random graphs sampled from the recently introduced \cite{newman2014equitable} statistical ensemble of equitable graphs, random graphs with a regular block structure. 
Following the derivation in \cite{rogers2008cavity} a finite set of non-linear equations for the spectral density is found and the solutions are provided for core-periphery structures. \\
The paper is organized as follows: in Section \ref{sec:reg} the equitable graph ensemble is defined and a generative algorithm is introduced.
In Section \ref{sec:spec} a brief introduction of the cavity approach to the computation of the spectral density of random matrices is provided and the general expression for the cavity variances of equitable graphs is shown, as previously derived in \cite{newman2014equitable,barucca2017spectral}. 
In Section \ref{sec:coreperi} we derive the main result of the paper, that is the analytic expression for the spectral density of a large class of equitable core-periphery graphs.  
In Section \ref{sec:con} we discuss the perspectives given by the use of the cavity method for spectral analysis of random matrices and for the detailed analytic description of the spectral properties of known graph structures. 

\section{\label{sec:reg}Equitable graphs}
A statistical ensemble of equitable graphs is defined by a set of vertices $V$, a partition $B=\{B_a\}_{a=1}^m$ dividing $V$ in $m$ non-overlapping blocks of vertices and a connectivity matrix $\mathbf{c}$, a $m\times m$ matrix of non-negative integer numbers.
Each undirected graph $G=(V,\,E)$ of an ensemble of equitable graphs must satisfy the constraints:
\begin{equation}\label{eq:rrBMdef}
\forall a,b \in B\:\forall i \in a \, \, \,  |\{(i,j) \in E\,|\,j \in b \}| = c_{ab},
\end{equation}
i.e. the total number of edges of node $i$ in block $a$ with a vertex in $b$ equals $c_{ab}$, for every vertex $i$ and every pair of blocks $a$ and $b$. 
In the general case of blocks of different sizes, $|B_a|=N_a$, then, for the system (\ref{eq:rrBMdef}) to have solution, the connectivity matrix $\mathbf{c}$ and block sizes must obey the equitability condition: 
\begin{equation}\label{eq:rrBMcon}
\forall a,b \in B\:N_ac_{ab} = N_bc_{ba},
\end{equation}
i.e. the total number of edges between blocks $a$ and $b$ must be uniquely defined. 
All graphs satisfying (\ref{eq:rrBMdef}) have equal probability in the ensemble. \\
If we introduce the block degrees $k_{i\rightarrow a} =  |\{(i,j) \in E | j \in a \}| $, (\ref{eq:rrBMdef}) can be reformulated as follows: the vector of block degrees of each node in a given block equals the row of the connectivity matrix corresponding to the block index, i.e. $\forall i \in a\:k_{i\rightarrow b} = c_{ab}$. 
Equitable graphs represent a block structured generalization of k-regular random graphs in the sense that when $c_{ab} = c $ for all pairs $(a,b)$ the corresponding blockmodel ensemble is the set of k-regular random graphs with $k=cm$.\\ 
The form of the regularity constraints, Eq.(\ref{eq:rrBMdef}), allow edges to be drawn independently for each pair of blocks, and in particular, for the case of blocks of the same-size, it is possible to sample equitable graphs simply by assembling regular graphs \cite{wormald1999models}. 
Between each pair of blocks the edges are drawn according to a k-regular graph, where the value of $k$ equals the corresponding element of the connectivity matrix, then the total set of edges is given by the union of the sets for each of the $m^2$ regular and biregular graphs [Algorithm \ref{alg:eqgraphs}].  \\
This ensemble has been successfully studied in the context of community detection \cite{brito2016recovery} and an algorithm was found that is able to identify blocks simply by looking at the list of edges in the graph \cite{barucca2017spectral}, exploiting the symmetry of the eigenvectors of the adjacency matrix. 

\begin{algorithm}
    \label{alg:eqgraphs}
   \caption{Equitable Graphs}
    \begin{algorithmic}[1]
      \Function{Generate Equitable Graph}{$N_a,c_{ab},m$}\Comment{Where N - array, c - matrix, m - integer. Outputs the lists of vertices and edges of an undirected equitable graph.}
        \If{Equation \eqref{eq:rrBMcon} applies}
            \For{$a = 1$ to ${m}$}
                \State Generate set of vertices $V_a$, such $|V_a|=N_a$
            \EndFor
            \For{$a = 1$ to ${m}$}
                \For{$b = a$ to ${m}$}   
                    \If{a == b}
                        \State $E_{aa}$ = Generate k-regular graph($k=c_{aa}$,$V_a$)
                    \Else
                       \State $E_{ab}$ = Generate k-biregular graph($k=c_{ab}$,$V_a$,$V_b$)
                    \EndIf
                \EndFor
            \EndFor
            \State $V = \cup_{a}^m V_{a}$
            \State $E = \cup_{a,b=1}^m E_{ab}$
        \Else
        \State Print 'Equitability condition does not apply'
        \EndIf
      \EndFunction

      \Function{Generate k-regular graph}{$k$,$V$}\EndFunction
      Where k - integer, V - set of vertices. Outputs the list of edges of an undirected regular graph between the vertices $V$ \cite{wormald1999models}. 
      \Function{Generate k-biregular graph}{$k$,$V_1$,$V_2$}\EndFunction
      Where k - integer, $V_{1/2}$ - sets of vertices. Outputs the list of edges of a bipartite regular graph from vertices $V_1$ to vertices $V_2$ \cite{wormald1999models}.
\end{algorithmic}
\end{algorithm}

\begin{figure}
\centering
\includegraphics[width=100mm]{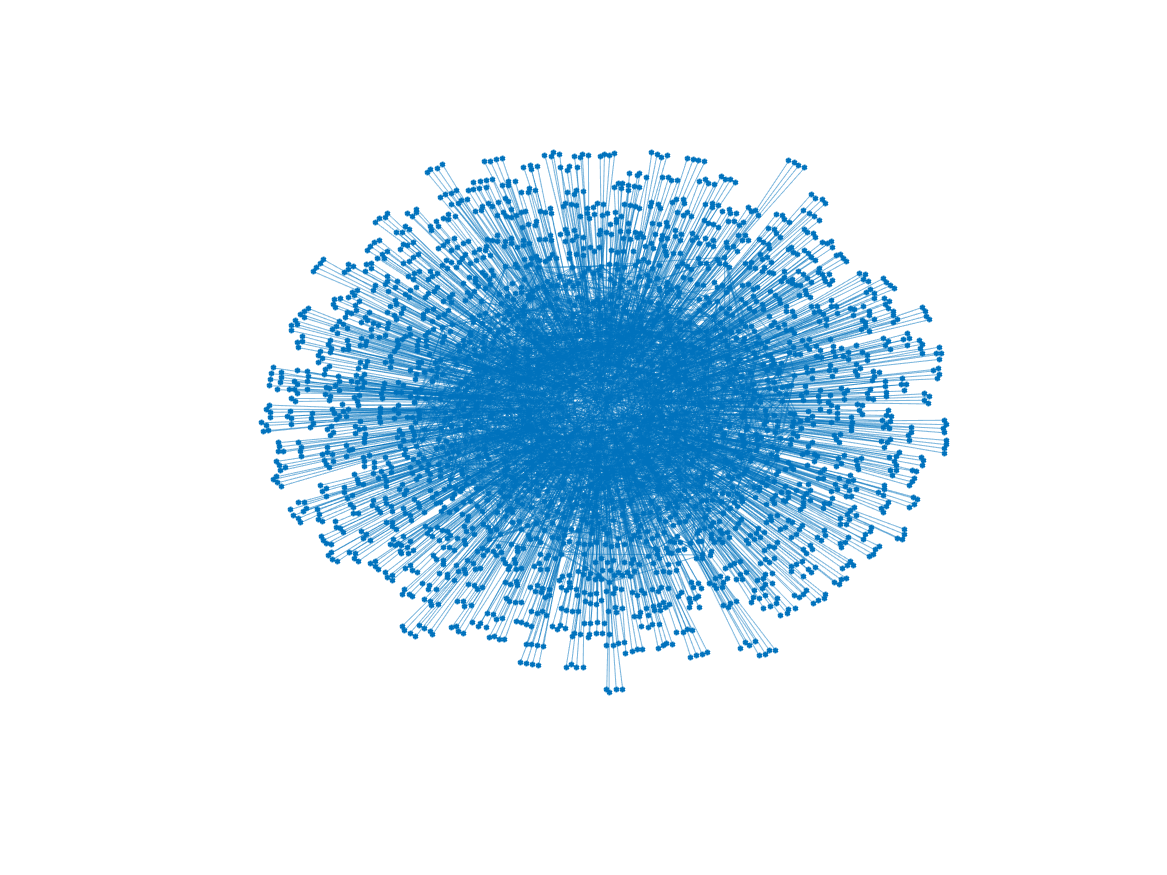} \protect\caption{Equitable core-periphery graph with $N_c =500$, $N_p=2000$, $c_{cc}=10$, $c_{cp}=4$, $c_{pc} =1$, and $c_{pp}=0$. Given the different sizes of the core and the periphery, $c_{cp}$ and $c_{pc}$ differ and their ratio is constrained by the relation $\frac{c_{cp}}{c_{pc}} = \frac{N_{p}}{N_c} = 4$. }
\label{fig:cpDOE} 
\end{figure}

\section{\label{sec:spec}Spectral theory with the cavity method}

We derive the cavity equations \cite{mezard1987spin, mezard2001bethe} to compute the continuous part of the spectrum of the adjacency matrices in equitable graphs, as shown in \cite{kuhn2011spectra,barucca2017spectral}.
Given an ensemble of $N\times N$ symmetric matrices the set of eigenvalues of a given adjacency matrix $A$ is denoted by $\{\lambda_i^A\}_{i=1}^N$. The corresponding empirical spectral density is defined as:
\begin{equation}\label{eq:specDen}
\rho(\lambda;A)=\frac{1}{N}\sum_{i=1}^N\delta(\lambda-\lambda_i^A),
\end{equation}
which satisfies the identity\cite{edwards1976eigenvalue}:
\begin{equation}\label{eq:edwards}
\rho(\lambda;A)=\frac{2}{\pi}\lim_{\epsilon\rightarrow 0^+}\frac{1}{N}\Im\left[\frac{\partial}{\partial z}\log\mathcal{Z}(z;A)\right]_{z=\lambda-i\epsilon}
\end{equation}
where $\Im[z]$ stands for  the imaginary part of the complex number $z$ and where $\log\mathcal{Z}(z;A)$ can be expressed via Gaussian integrals as in\cite{edwards1976eigenvalue}, i.e.: 
\begin{equation}\label{eq:Z}
\mathcal{Z}(z;A)=\int \left[\Pi_{i=1}^N\frac{dx_i}{\sqrt(2\pi)} e^{-H(x;z,A)}\right]
\end{equation}
with $H(x;z,A) = \frac{z}{2}\sum_{i}^N x_i^2  -\frac{1}{2}\sum_{i,j}^N A_{ij}x_ix_j $. From this formula, an expression for the spectral density of any graph of the ensemble can be derived in terms of the variances of the Gaussian variables introduced in (\ref{eq:Z}), 
\begin{equation}\label{eq:specDensVar}
\rho(\lambda;A)=\frac{1}{\pi}\lim_{\epsilon\rightarrow 0^+}\frac{1}{N}\Im\left[\sum_i^N\langle x_i^2\rangle_z\right]_{z=\lambda-i\epsilon}.
\end{equation}
Finding the variances in (\ref{eq:specDensVar}) is not generally more straightforward than numerically diagonalizing the matrix $A$ but an approximation method has been introduced for sparse graphs that holds exactly in the large $N$ limit, the cavity method \cite{mezard2001bethe}.\\
Briefly, conditional probability distributions are introduced for each node and are parametrized by specific variables, i.e. the cavity variances $\Delta_i^{(j)}$, each representing the variance of $x_i$ if its neighbor $j$ is not taken into account.
With such approximation the following set of self-consistent equations can be derived \cite{rogers2008cavity}:
\begin{equation}\label{eq:cavityVar}
\Delta_{i}^{(j)}(z)=\frac{1}{z-\underset{l\in\partial i\setminus j}{\overset{N}{\sum}}A_{il}^{2}\Delta_{l}^{(i)}(z)},
\end{equation}
where $\partial i$ is the set of neighbor of node $i$, i.e. $\partial i = \{e \in E | i \in e\}$.
From cavity variances it is possible to compute node variances via the equations
\begin{equation}\label{eq:nodeVar}
\Delta_{i}(z)=\frac{1}{z-\underset{l\in\partial i}{\overset{N}{\sum}}A_{il}^{2}\Delta_{l}^{(i)}(z)},
\end{equation}
which lead to compute the spectral density $\rho(\lambda;A)$. \\
In the case of equitable graphs, the ansatz of \textit{block-simmetry} can be made for the cavity variances:
\begin{equation}\label{eq:ansatz}
\Delta_{i}^{(j)}(z) = \Delta_{g_i}^{(g_j)}(z).
\end{equation}
This exact ansatz, also described in \cite{newman2014equitable,barucca2017spectral} reduces the set of equations for cavity variances from a size of order $N$ (in the sparse case) to a set of $m^2$ equations:
\begin{equation}\label{eq:ansatzCav}
\Delta_{a}^{(b)}(z)=\frac{1}{z-\underset{c}{\overset{m}{\sum}}(c_{ac}-\delta_{bc})^+\Delta_{c}^{(a)}(z)},
\end{equation}
where $(x)^+ = \max(x,0)$.
Then, block variances can be obtained,
\begin{equation}\label{eq:blockVar}
\Delta_{a}(z)=\frac{1}{z-\underset{c}{\overset{m}{\sum}}c_{ac}\Delta_{c}^{(a)}(z)},
\end{equation}
and eventually we get a simplified formula for the spectral density,
\begin{equation}\label{eq:rrBMspec}
\rho(\lambda)=\frac{1}{\pi N}\underset{a=1}{\overset{m}{\sum}}N_a\Im[\Delta_{a}(z)]_{z=\lambda-i\epsilon},
\end{equation}
which is now restricted to a weighted sum of the block variances. 

\section{\label{sec:coreperi}Spectral theory of equitable core-periphery graphs}
In the following we will consider equitable core-periphery graphs with a set of core nodes of size $N_c$, a set of periphery nodes of size $N_p$ and a connectivity matrix $\mathbf{c}$, 
\begin{equation}
c=\left(\begin{array}{ccc}
c_{cc} & c_{cp}\\
c_{pc} & c_{pp}
\end{array}\right),\label{eq:affmat}
\end{equation}
where the only constraint we here assume is that $c_{cc}+c_{cp}>c_{pc} + c_{pp}$, i.e. the degree of a node in the core is stricly larger than the degree of node in the periphery.
The degree regularity in equitable core-periphery graphs trivializes the problem of identifying which nodes are part of the core and which are part of the periphery, since degree alone is sufficient to establish unambiguously the assignment of a node. 
A feature which partially characterizes also the corresponding problem in the stochastic block model core-periphery case \cite{zhang2015identification}. \\ 
Here is shown how isolated eigenvectors of the adjacency matrix entail exact information on the block structure of equitable graphs. 
Starting from the secular equation
\begin{equation}\label{eq:secEqA}
\underset{j=1}{\overset{N}{\sum}}A_{ij}u_{j}=\lambda u_{i},
\end{equation}
an ansatz of \textit{block-simmetry} can be made such that nodes in the same block share the same eigencomponent, i.e. for all $i$ is hypothesized that $u_i = u_{g_i}$. 
Since the number of neighbors between different groups is fixed it follows that:
\begin{equation}\label{eq:secEqC}
\underset{b=1}{\overset{m}{\sum}}c_{ab}N_bu_{b}=\lambda u_{a},
\end{equation}
which yields the useful conclusion that each block-symmetric eigenvector of the adjacency matrix corresponds to an eigenvector of the block-size weighted connectivity matrix $\tilde{c}_{ab}= c_{ab}N_b$, and viceversa. 
These eigenvectors correspond to a finite set of non-densely distributed eigenvalues.

Let us now consider a specific choice of the parameters, 
\begin{equation}
c=\left(\begin{array}{ccc}
k & k'\\
1 & 0
\end{array}\right),\label{eq:affmat}
\end{equation}
and, as prescribed by the equitability condition, $N_p/N_c = k'$. \\
In this class of equitable core-periphery graphs, the core sub-graph constitutes a k-regular graph, the connections between the $N_c$ nodes in the core and the $N_p$ nodes in the periphery define a bi-regular graph of core-periphery degree $k'$, i.e. each node in the core has exactly $k'$ neighbors in the periphery and, on the other hand, each node in the periphery has exactly one neighbor in the core. \\
Further, the class of core-periphery graphs considered, without any internal links within the periphery, yields a number of zero eigenvalues equal to the difference between the periphery size and the core size. \\
This can be readily demonstrated by considering the block structure of the adjacency matrix, 
\begin{equation}
A=\left(\begin{array}{ccc}
A^{cc} & A^{cp}\\
A^{pc} & A^{pp}
\end{array}\right),\label{eq:affmat}
\end{equation} where each sub-block represents the adjacency matrix of the corresponding sub-graph. 
In particular, $A^{pp}_{ij}= 0$ for each pairs of nodes in the periphery. From this property and from the regularity conditions, it follows that:
\begin{align*}\label{eq:zeroeigs}
\det(A - \lambda I_{N_c+N_p}) &= \det(A^{cc} - \lambda I_{N_c} - \lambda^{-1}A^{cp} I_{N_p} A^{pc} )\det(\lambda I_{N_p}) \\
&=\det(\lambda A^{cc} - (\lambda^2 + k')I_{N_c})\lambda^{N_p - N_c},
\end{align*}
where $I_n$ indicates the identity matrix of size $n$.
The block-regular structure also allows us to compute the two eigenvalues corresponding to fully block-symmetric eigenvectors, Eq.\ref{eq:secEqC}, which read $\lambda_{\pm} = \frac{k}{2} \pm \sqrt{(\frac{k}{2})^2 + k'}$.
\\
We now turn to the main computation of the continuous part of the spectrum with the cavity method. 
The intrinsic block heterogeneity of core-periphery structure makes it incompatible with the Ansatz of full symmetry that can be used for modular and bipartite structures \cite{barucca2017spectral}. 
In particular, the cavity variances for nodes in the core and nodes in the periphery clearly differ:
\begin{align} 
\Delta_{c}^{(c)}(z)&=\frac{1}{z-(k-1)\Delta_{c}^{(c)}(z)-k'\Delta_{p}^{(c)}(z)} \\
\Delta_{c}^{(p)}(z)&=\frac{1}{z-k\Delta_{c}^{(c)}(z) - (k'-1)\Delta_{p}^{(c)}(z) } \\
\Delta_{p}^{(c)}(z)&=\frac{1}{z} \\
\Delta_{p}^{(p)}(z)&=\frac{1}{z-\Delta_{c}^{(p)}(z)}.
\end{align}
Thanks to the choice of the parameters for this class of equitable core-periphery graphs, the only equation to be solved in the system of cavity equations turns out to be 
\begin{align*} 
\Delta_{c}^{(c)}(z)&=\frac{1}{z-\frac{k'}{z}-(k-1)\Delta_{c}^{(c)}(z)},
\end{align*}
that yields the solution:
\begin{align} 
   \alpha \equiv \Re[\Delta_{c}^{(c)}(z)]_{z=\lambda-i0^+}&=\begin{cases}
        \frac{\mu\pm\sqrt{\mu^2-4(k-1)}}{2(k-1)}, & \text{if}\:\:|\mu|\geq2\sqrt{k-1}\\
        \frac{\mu}{2(k-1)}, & \text{otherwise}
        \end{cases} \\
   \beta \equiv \Im[\Delta_{c}^{(c)}(z)]_{z=\lambda-i0^+}&=\begin{cases}
        0, &\text{if}\:\:|\mu|\geq2\sqrt{k-1}\\
        \frac{\sqrt{4(k-1)-\mu^2}}{2(k-1)}, & \text{otherwise},
        \end{cases}
\end{align}
where $\mu=\lambda - \frac{k'}{\lambda}$. 
Since a non-zero imaginary part is a necessary condition for a non-zero support of the spectral density, we can immediately find the boundaries of its support from the condition $|\lambda-\frac{k'}{\lambda}|=2\sqrt{k-1}$. \\
From $\Delta_{c}^{(c)}(z)$, can then be derived $\Delta_{c}^{(p)}(z)$, $\Delta_{p}^{(p)}(z)$, and finally the block variances $\Delta_{c/p}$:
\begin{align} 
   [\Delta_{c}(z)]_{z=\lambda-i0^+}&= \frac{1}{\mu - k(\alpha + i\beta)} \\
   [\Delta_{p}(z)]_{z=\lambda-i0^+}&= \frac{1}{\lambda} \frac{\mu +1/\lambda -k(\alpha + i\beta)}{\mu - k(\alpha + i\beta)},
\end{align}
which yield the analytic form of the spectral density:
\begin{equation}\label{eq:specDensCP}
\rho(\lambda) = \frac{k\beta}{2\pi\delta}\left(   1 + \frac{k'}{\lambda^2} \right), \:\:\:\: \:\:\:\text{with}\:\:|\lambda-\frac{k'}{\lambda}|<2\sqrt{k-1},
\end{equation}
where $\delta = (\mu-k\alpha)^2+(k\beta)^2$. More explicitly: 

\begin{equation}\label{eq:specDensCP}
\rho(\lambda) = \frac{k}{2 \pi}\frac{\sqrt{(k-1)-\frac{1}{4}\left(\lambda-\frac{k'}{\lambda}\right)^2}}{k^2-\left(\lambda-\frac{k'}{\lambda}\right)^2}  \left(   1 + \frac{k'}{\lambda^2} \right), \:\:\:\: \:\:\:\text{with}\:\:|\lambda-\frac{k'}{\lambda}|<2\sqrt{k-1},
\end{equation}
where the normalization factor presented is such that the density integrates to one when restricted to the non-zero eigenvalues.

\begin{figure}
\centering
\includegraphics[width=100mm]{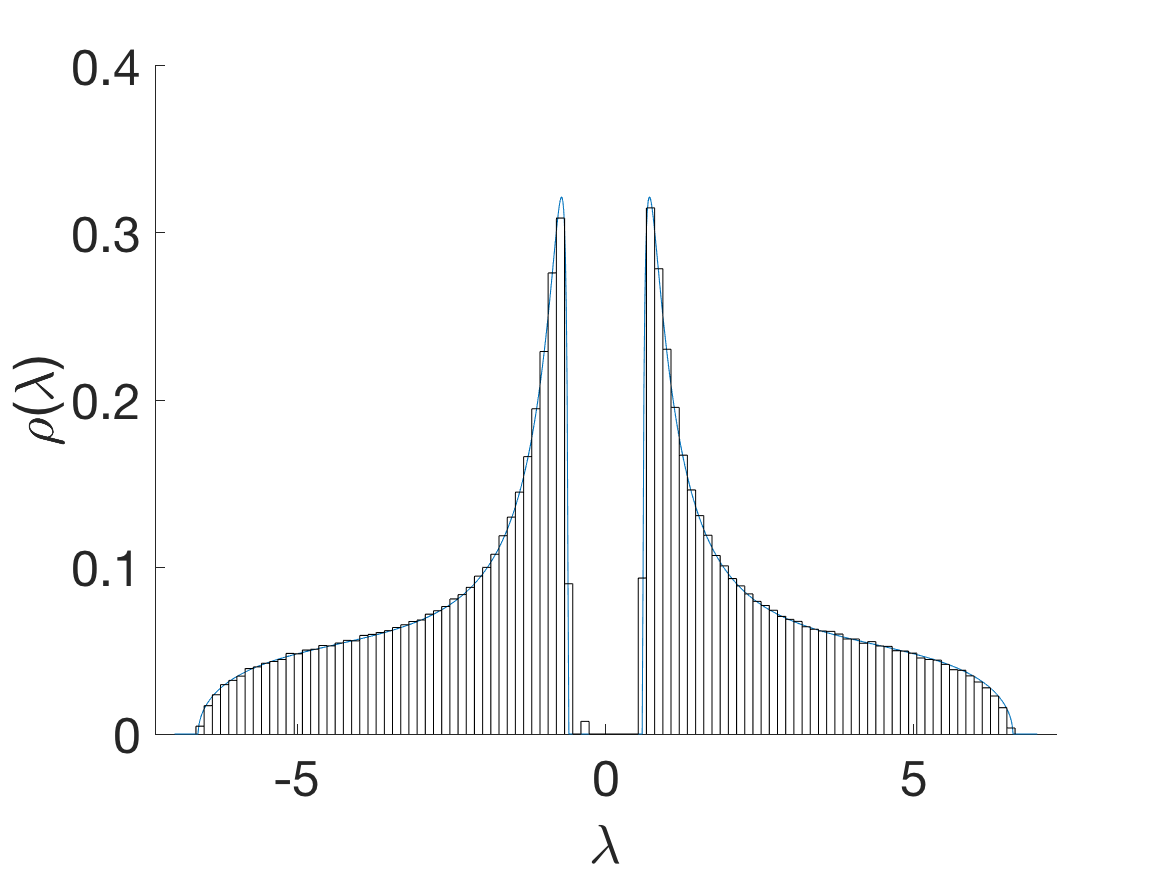} \protect\caption{Spectral density Equitable graph for $k=10$, $k'=4$ corresponding to Eq.(\eqref{eq:specDensCP}). Bars come from numerical diagonalization of a sample of $100$ equitable graphs of total size $N=2500$. In the numerical diagonalisation, for the interval considered, we recognise the eigenvalue $\lambda_{-}=\frac{k}{2}-\sqrt{ (\frac{k}{2})^2 + k'}$, while the top eigenvalues $\lambda_{+}$ falls outside the range of the plot. }
\label{fig:cpDOE} 
\end{figure}

\section{\label{sec:con}Conclusions}
In this paper, via the cavity method introduced in \cite{mezard1987spin, mezard2001bethe}, a new class of spectral laws has been explicitly derived for equitable core-periphery graphs and naturally generalizes the long-known Kesten-McKay law for k-regular graphs \cite{kesten1959symmetric,mckay1981expected}. \\ 
This analytic result provides a general benchmark to easily compare spectra of real and artificial arbitrarily large networks presenting a core-periphery structure with the spectrum of an idealized equitable case.

\bibliographystyle{model1-num-names}

\bibliography{sample}

\end{document}